# URIUM: A Programming Language for a Practical Open Course on Compiler Design

Francisco J. MORENO-VELO[1*], Almudena GARCÍA JURADO-CENTURIÓN[2]
[1] *Departamento de Tecnologías de la Información, Universidad de Huelva, Huelva, Spain*
[2] *Departamento de Tecnologías de la Información, Universidad de Huelva, Huelva, Spain*
*e-mail: francisco.moreno@dti.uhu.es, almudena.garcia@dti.uhu.es*

**Abstract.** This paper presents the definition of a simple programming language used as the basis for developing a practical compiler design course. The course explains step by step how to build a compiler, from the initial analysis stages to code generation. The developed compiler generates code for various processors (MIPS, Intel, and RISC-V) and operating systems (MS-Windows and Linux). The course can be adapted to different levels of difficulty and can be used as a starting point for explaining more advanced topics.

**Author's Note:** The work presented here is an open course. All course content, including the source code for each lecture, is available at the following link. The instructions for each session are included in the repository's wiki. https://github.com/fran-moreno-velo/urium

## 1. Introduction

The study of compilers is considered important in the curriculum for Computer Science graduates for several reasons. First, compiler design includes the development of lexical, syntactic, and semantic analyzers, and these analysis techniques are very useful in many types of programs. Second, compilers are highly complex programs, so their design provides an excellent practical application of Software Engineering techniques. Third, developing a compiler requires detailed knowledge of operating systems and computer architecture, since the object code generated by a compiler depends substantially on the target processor and the operating system it uses. Finally, compiler design requires a deep understanding of the programming language to be compiled, so studying it allows students to consolidate and deepen their programming knowledge.

In compiler design courses, it's common to use the code of a particular compiler as teaching support material. This is usually a compiler for a very basic programming language, both in its instructions and data types, as well as in the final code generated, which is typically intermediate code for a virtual machine. A widely used example is the PL/0 language, proposed by Wirth (1976). Some universities are opting for slightly more complex languages that contain object-oriented features, such as the Cool language from University of California, Berkeley (Aiken (1996)), the miniJava language proposed by Roberts (Roberts (2001)), or Wirth's Oberon-0 language (Wirth (2017)). This paper presents a new programming language proposed as the basis for a practical compiler design course. The course content consists of developing a basic compiler that generates assembly code for real processors and operating systems. This approach allows for a much deeper exploration of computer architecture and operating system aspects that are covered in less detail in many other compiler design courses.

## 2. Course structure

This practical course on Compiler Design uses a simple programming language called URIUM as a case study to develop the different stages of the compilation process. The course consists of 12 sessions that can be adapted to cover a semester. The course syllabus is as follows:

- Lecture 00: The URIUM programming language
- Lecture 01: Lexical analyzer
- Lecture 02: Syntactic analyzer

---

* Corresponding author

- Lecture 03: The Abstract Syntax Tree
- Lecture 04: Semantic analysis - Header parser
- Lecture 05: Semantic analysis - Body parser
- Lecture 06: Using an automatic tool
- Lecture 07: Intermediate code generation
- Lecture 08: Backend for MIPS32 emulator
- Lecture 09: Backend for MS-Windows on Intel64
- Lecture 10: Backend for Linux on AMD64
- Lecture 11: Backend for RISC-V simulator
- Lecture 12: Backend for Linux on RISC-V

The course begin with the definition of the language and progress to building a compiler that works on different platforms. The first sessions are dedicated to the compiler analysis phase. This includes manually programming the language's lexical analyzer (Lecture01), manually programming the language's syntactic analyzer (Lecture02), manually programming the language's semantic analyzer (Lecture03, Lecture04, and Lecture05), and a version of the analyzers automatically generated by a parser generator tool (Lecture06).

Special attention has been given to the compiler's backend, generating object code for different processors (MIPS, Intel, RISC-V) and operating systems (MS-Windows, Linux). The second part of the course begins with intermediate code generation (Lecture 07). Subsequent sessions focus on object code generation for the various platforms. Lecture08 describes code generation for the simulated MIPS32 processor using Qt-Spim. Lecture09 describes code generation for the MSWindows operating system on the Intel64 architecture. Lecture10 focuses on code generation for the Linux operating system on the AMD64 architecture. Lecture11 describes code generation for the simulated RISC-V processor using RARS. Lecture12 focuses on code generation for the Linux operating system on the RISC-V architecture.

## 3. The URIUM programming language

### *3.1. Language definition*

URIUM is the name the Romans gave to the Tinto River, located in the province of Huelva, Spain. This river is known for the reddish color of its waters, which is due to heavy metal sediments. The river originates in a region that has been mined since ancient times by the Iberians, Phoenicians, Romans, and Muslims. Copper and pyrite (iron sulfide) have been the main minerals extracted from its mines. Numerous archaeological remains have been found along the river, demonstrating the importance of this area since antiquity. In 1930, a well-preserved Corinthian helmet from the 6$^{th}$ century BC was discovered at its mouth. This helmet is one of the most important archaeological pieces from that period found in Spain and is on display at the National Archaeological Museum in Madrid.

The name URIUM has been chosen for the programming language because rivers have historically been very important communication channels and this river runs through the entire province of Huelva. For this reason, the language logo is a Greek helmet, and the language brand is described in capital letters using a Roman font. Fig. 1 shows the image used as the header in the course documents, where the logo and brand appear on a marble-textured background.

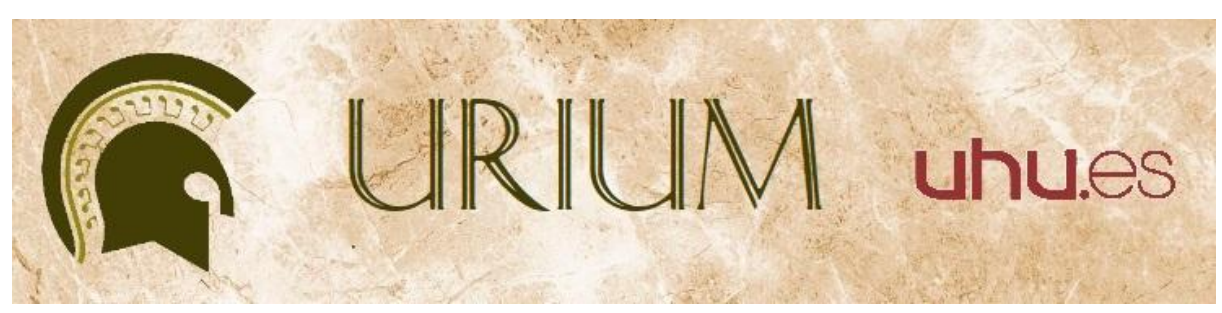

Fig. 1. The URIUM logo and brand on a marble textured background.

URIUM is a process-oriented, imperative programming language (similar to C). The goal of URIUM is to define a very simple programming language to support the teaching of a basic course on Compiler Design. In each aspect of URIUM's design (lexical, syntactic, and semantic specifications, data types, instructions, etc.), the language has been simplified to allow for compiler development both manually and with the aid of tools. In this way, the URIUM language will serve as an example in the practical sessions that explain how to develop the different stages of a compiler. Fig. 2 shows an example of the contents of a source file described in URIUM language.

Some of the features of URIUM are the following:

- **Reserved words**:

  URIUM includes a small number of reserved words (17 in total). Compared to C (32 reserved words), C++ (95 reserved words), or Java (68 reserved words), the simplification effort made in the language design is noteworthy. The reserved words in URIUM are as follows: *as*, *boolean*, *char*, *else*, *endp*, *false*, *if*, *import*, *int*, *library*, *native*, *out*, *private*, *proc*, *public*, *true*, and *while*.
- **Data types**:

  The language includes only three data types: *int*, *char*, and *boolean*. It is assumed that all of them will be stored in 32-bit registers. The *char* type is considered to be described in 8-bit ASCII code stored in the least significant byte (the 24 most significant bits must have the value 0). The *boolean* type uses the value 0 to indicate false and the value 1 to indicate true.
- **Statements**:

  URIUM procedures consist of a list of statements. The statements included in the language are: declarations of local variables (assignment at the time of declaration and joint declaration of a list of variables of the same type are supported), simple assignment instructions (using the = operator), statements for executing a procedure, conditional statements (*if-else*), loops (*while*), and procedure termination (*endp*).
- **Arithmetic and logical operators**:

  URIUM includes the following arithmetic operators: addition (+), subtraction (−), multiplication (∗), division (/), and remainder (%).

  For their part, the logical operators included in URIUM are the following: and with short circuit evaluation (&&), or with short circuit evaluation (| |) and not (!).
- **Relational operators**:

  Relational operators allowyou to compare the values of different expressions. URIUMincludes four relational operators that can be applied to int and char expressions: greater than (>), less than (<), greater than or equal to (>=), and less than or equal to (<=). It also includes the equal (==) and not equal (! =) operators, which are valid for all data types (int, char, and boolean).
- **Literals**:

  URIUM includes literals for each data type: *int* values (in decimal, octal, hexadecimal, and binary format), *char* values (in single quotes that can include printable characters, escape characters, and characters in octal format), and *boolean* values (*true* or *false*).
- **Libraries and procedures**:

  Files described in the URIUM language use the “*.ur*” extension and define procedure libraries. Each file contains a list of imported libraries (*import* clauses) and a *library* whose name must match the filename. A library name can consist of identifiers separated by periods. In that case, the identifiers indicate the path where the source file should be stored. For example, a library called *mypkg.mysubpkg.MyLib* should be located at “*/mypkg/mysubpkg/MyLib.ur*”.

  The library consists of a list of procedures, which can be *public* or *private*. The code of the procedures can contain calls to other public or private procedures within the same library. They can also contain calls to public procedures from other libraries. In the latter case, the called procedure is identified by the library name, followed by a period and the procedure name. To use the procedures of a library, it must have been previously imported into the file (an alias can be assigned to the library at this point).

  The language also allows the definition of *native* libraries, which consist of a list of procedure declarations. In this case, the procedure code is written directly in assembly language and is not included in the source file. Thanks to these native libraries, it is possible to use procedures that use specific assembler instructions that are not generated by the compiler, such as system calls.
- **Applications**:

  To develop an application in URIUM, you must define a library called *Main* that must contain a public procedure called *main()*, which represents the entry point of execution. The application can consist of many other auxiliary libraries that are incorporated using import clauses.

```
import urium.Console as Console;
/* Application that shows the Greatest Common Divisor of two numbers */
library Main
{
	/* Application start procedure */
	public proc main()
	{
		int a=54;
		int b=24;
		int c;
		gcd(a, b, out c);
		Console.print('G');
		Console.print('C');
		Console.print('D');
		Console.print('(');
		Console.print(a);
		Console.print(',');
		Console.print(b);
		Console.print(')');
		Console.print('=');
		Console.print(c);
	}
	/* Computes the Greatest Common Divisor of two numbers */
	private proc gcd(int a, int b, out int result)
	{
		int gcd = a;
		if (b<gcd) gcd = b ;
		while (gcd>0)
		{
			if ( a%gcd == 0 && b%gcd == 0)
			{
				result = gcd;
				endp;
			}
			gcd = gcd - 1;
		}
		result = 1;
	}
}
```

Fig. 2. Example of a source file described in URIUM language.

### *3.2. Native libraries*

The URIUM language allows the inclusion of libraries declared as native. These libraries contain a set of precompiled procedures. This means that the URIUM compiler will not generate the assembly code for these procedures but will simply copy the existing code. This allows customization of the code for these procedures, including system calls that the compiler does not generate automatically. The URIUM distribution includes two native libraries (**urium.Console** and **urium.Program**) with their corresponding assembly code for the different platforms supported by the compiler.

The **urium.Console** library includes procedures dedicated to managing data input and output to the text console. These procedures allow you to display data in different formats (characters, decimal integers, binary integers, hexadecimal integers) and read character and integer data (Fig. 3).

```
/* Library of procedures for accessing the console */
native urium.Console
{
	/* Writes a character to the console */
	public proc print(char c);
	/* Writes an integer to the console */
	public proc print(int i);
	/* Writes an integer to the console in binary format */
	public proc printBits(int i);
	/* Writes an integer to the console in hexadecimal format */
	public proc printHex(int i);
	/* Reads an integer from the console */
	public proc readInt(out int i);
	/* Read a character from the console */
	public proc readChar(out char c);
}
```

Fig. 3. Source code for the urium.Console native library.

The native library **urium.Program** contains procedures for accessing application call arguments. These procedures allow you to determine the number of arguments entered on the command line and access their values, whether integers or strings. Since URIUM does not support the string data type, the library provides a procedure for accessing strings that obtains the argument length and another that allows you to access each character of that argument. The library also contains the *exit()* procedure to force the application to terminate (Fig. 4).

```
/* Library of procedures for accessing application arguments */
native urium.Program
{
	/* Gets the number of arguments */
	public proc getArgCount(out int argc);
	/* Gets the length of the i-th argument */
	public proc getArgLength(int index, out int length);
	/* Gets the i-th argument as an integer */
	public proc getIntArg(int index, out int arg);
	/* Gets the j-th character of the i-th argument */
	public proc getCharArg(int index, int pos, out char arg);
	/* Ends the program execution */
	public proc exit(int code);
}
```

Fig. 4. Source code for the urium.Program native library.

### *3.3. The URIUM compiler*

The URIUM compiler is located in the *uriumc.jar* file. To perform a compilation process, this file must be executed from a command line:

**java -jar uriumc.jar options**

The compiler's objective is to compile the *Main.ur* file, which must contain the *main()* procedure that starts the programmed application. The remaining imported libraries are then analyzed and compiled from this main library. By default, the working directory is the directory where the compilation command is executed, but this can be changed using the options. Imported libraries are also located in the working directory by default, but other search directories can be configured using the -I option. The code for precompiled libraries is searched for in the directories defined by the -L option. The compilation result is a file with the "*.s*" extension containing the assembly code that develops the programmed application. By default, the output file is named "*Application.s*", but this name can be changed using the -o option. The -v option can be used to instruct the compiler to generate files with a description of the intermediate code in text format. The platform for which the code is generated (the compiler's backend) is selected with specific options. The compiler supports the following options:

- **Path**, indicates the working directory.
- **-o Name**,indicates the name of the output file (without the ".s" extension).
- **-I Path**, add a directory for searching imported files.
- **-L Path**, adds a directory for searching pre-compiled libraries.
- **-v**, indicates that intermediate code files should be generated (by default they will not be
- generated).
- **-none_mips32**, choose the backend on the MIPS32 processor simulated with Qt-Spim.
- **-win_x64**, choose the backend for the MS-Windows operating system on the Intel64 architecture.
- **-linux_amd64**, choose the backend for the Linux operating system on the AMD64 architecture.
- **-none_riscv**, choose the backend for the RISC-V processor simulated with RARS.
- **-linux_riscv**, choose the backend for the Linux operating system on a RISC-V board.

The URIUM distribution includes several examples, an "*include*" directory with the native library definitions, several directories with precompiled code for the native libraries on different platforms, and a "*Compilation_scripts*" directory containing the compiler and versions of compilation, assembly, and linking scripts for the various supported platforms. To compile an example, the easiest way is to copy the compiler and compilation scripts for the desired platform into the example directory and run these scripts. On emulated platforms (MIPS32 and RISC-V), the result is an assembler file that can be loaded and executed in the corresponding simulator (Qt-Spim, RARS). On the MSWindows platform, the generated file is assembled and linked using the *ml64* assembler distributed by Microsoft. On the Linux operating system, the generated file is assembled with the *as* command and linked with the *ld* command to obtain the executable.

## 4. The analysis phases on the compiler design

### *4.1. The lexical analyzer*

The first step in building the compiler is to develop the language's lexical analyzer. Lexical description is usually done by defining each lexical category using regular expressions. From these definitions, a deterministic finite automaton can be constructed whose final states correspond to the language's lexical categories. The process of generating the DFA from regular expressions should be explained in the theoretical part of the course and can be found in numerous manuals. Fig. 5 shows a portion of the DFA related with the comments and whitespace in the URIUM language. Lecture 01 explains how to manually program a lexical analyzer from a DFA and introduces the code for the URIUM compiler's lexical analyzer. During this lecture, students can be assigned exercises involving the introduction of new lexical categories into the analyzer.

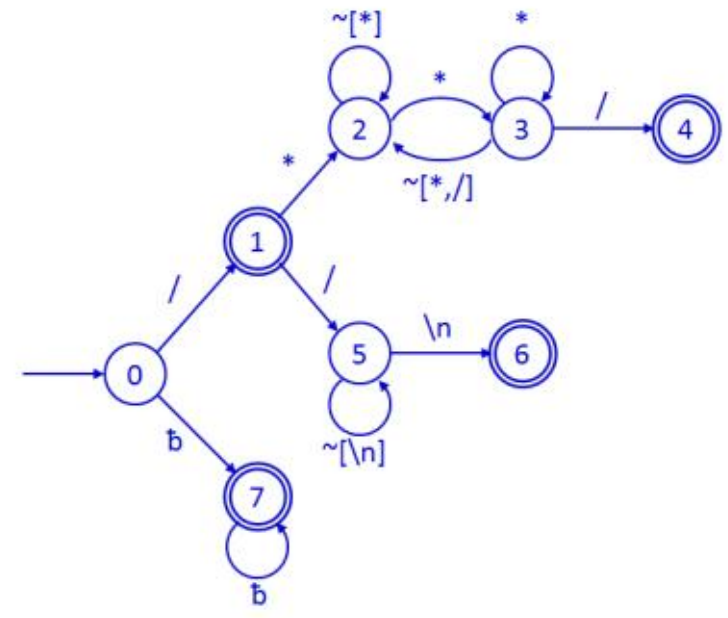

Fig. 5. A snippet from the URIUM lexical analyzer's DFA related to comments and whitespace.

*4.2. The syntactic analyzer*

The next step in compiler construction is developing the syntactic analyzer. This module verifies that the token stream generated by the lexical analyzer corresponds to a specific context-free grammar. This course uses recursive top-down parsing to develop this module. The URIUM language grammar, written in EBNF format, is presented in this phase. The theoretical sessions of the course should explain how to transform grammars written in EBNF format into grammars written in BNF format and how to calculate the prediction sets of the BNF rules necessary to implement top-down parsing. This content can be found in numerous manuals. Lecture 02 explains how to program the function related with recognizing a non-terminal symbol from the prediction sets of its BNF rules and develops the code for the complete URIUM grammar. Fig. 6 shows an example of a function that recognizes a symbol in recursive top-down parsing.

```
/*                                                     */
/* Parses the <A> symbol                               */
/* (Symbol <B> begins with TOKEN1, TOKEN2 or TOKEN3)   */
/*                                                     */
/* <A> ::= <B> <C> <D>                                 */
/* <A> ::= TOKEN4 <E> TOKEN5 <F>                       */
/*                                                     */
private void parseA() throws SintaxException
{
        int [] expected = { TOKEN1, TOKEN2, TOKEN3, TOKEN4 } ;
        switch (nextToken.getKind())
        {
                case TOKEN1:
                case TOKEN2:
                case TOKEN3:
                {
                        parseB();
                        parseC();
                        parseD();
                        break;
                }
                case TOKEN4:
                {
                        match(TOKEN4);
                        parseE();
                        match(TOKEN5);
                        parseF();
                        break;
                }
                default:
                {
                        throw new SintaxException(nextToken,expected);
                }
        }
}
```

Fig. 6. A function recognizing a non-terminal symbol in a top-down parser.

An important aspect of parsing is error handling. A compiler should not abort in the event of an error in the source file but should analyze the entire file, finding all errors and reporting them. Error handling in top-down parsing is particularly simple because it relies on the introduction of synchronization tokens, which are usually the tokens associated with separators and terminators (parentheses, braces, commas, and semicolons). Upon detecting a syntax error, the parser begins consuming tokens until it reaches a known position from which it can continue parsing. Fig. 7 shows the code for a function that includes error handling. This code uses the functions *catchError()* and *skipTo()* to add the detected error to the global error report and to consume the necessary tokens until a synchronization token is reached.

```
/*                                                                                    */
/* Parses the <A> symbol with error handling                                          */
/* (Symbol <A> ends with TOKEN10 and must be followed by TOKEN11 or  TOKEN12)         */
/*                                                                                    */
/* <A> ::= <B> <C> <D>                                                                */
/* <A> ::= TOKEN4 <E> TOKEN5 <F>                                                      */
/*                                                                                    */
private void parseA()
{
        int [] lsync = { TOKEN10 };
        int [] rsync = { TOKEN11, TOKEN12 };
        try
        {
                /* Code to parse the <A> symbol */
        }
        catch (Exception ex)
        {
                catchError(ex);
                skipTo(lsync,rsync);
        }
}
```

Fig. 7. A function recognizing a non-terminal symbol with error handling.

The URIUM grammar in BNF format has a total of 58 symbols and 108 rules. If a detailed explanation of the parser code is desired, Lecture02 can be divided into several sessions (for example, explaining error handling in a separate session). During this lecture, exercises can be carried out by proposing new syntactic rules and adding their analysis to the presented code.

### *4.3. The Abstract Syntax Tree*

The goal of a compiler's analysis phase is to construct a data structure that stores all the information contained in the analyzed source file. This structure is commonly known as an Abstract Syntax Tree (AST). It is a structure made up of objects that describe the language components (variables, expressions, statements, procedures, etc.). The semantic description of all the content of a source file is stored in the form of a tree where the root represents the most general component (a library in the case of URIUM) and the branches break down the information into more basic concepts (procedures, statements, etc.).
Lecture03 explains all the classes defined for developing the AST in the URIUM compiler. The session may include exercises in defining new structures to describe new possible functionalities of the language (new statements, for example). In this case, there are a total of 26 classes distributed across four packages as shown in Fig. 8. These packages are described below:

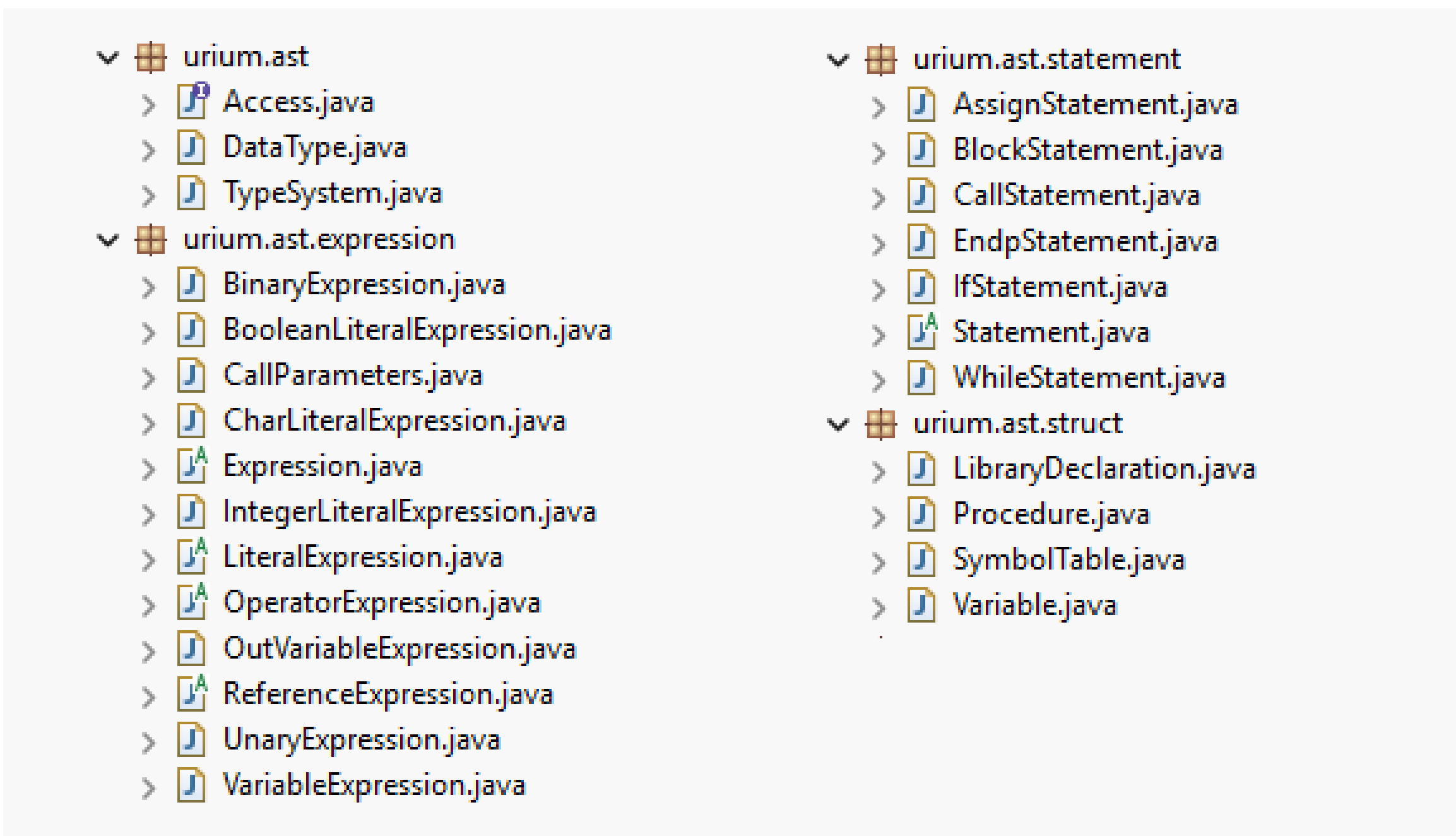

Fig. 8. Class structure used to define the Abstract Syntax Tree.

- **urium.ast**: It is the root package for all classes used in the data structure. The package only contains the classes that develop the type system, which are used by many of the classes in the various subpackages.
- **urium.ast.expression**: This package contains classes dedicated to describing the different arithmetic and logical expressions included in the URIUM language.
- **urium.ast.statement**: It is the package that includes the classes that describe the different statements of the URIUM language: (*if*, *while*, *endp*, etc.).
- **urium.ast.struct**: It is the package to which the classes that describe the high-level components of the URIUM language (libraries, procedures, etc.) belong.

### *4.4. The semantic analyzer*

The semantic analyzer is based on a modification of syntactic analysis in order to construct the AST while performing the analysis. This requires transforming the context-free grammars used in the syntactic definition into attributed grammars. Attributed grammars contain attributes associated with the symbols and semantic actions embedded in the rules. Converting top-down parsing into semantic analysis is straightforward, as the attributes are treated as arguments to the recognition functions for each symbol, and the semantic actions are directly included in the code of these functions. Fig. 9 shows an example of the code associated with recognizing the *WhileStm* symbol. In this case, the symbol uses the *symtab* attribute and constructs a **Statement** object that will be added to the AST.

The URIUM language does not use header files that allow predeclaring the contents of a library; in other words, there are no "*.h*" or "*.hpp*" files like those used in C or C++. This makes it impossible to compile a URIUM file in a single pass. The reason is that if a single pass is performed, the compiler only knows the contents of the part of the file that has already been analyzed, but cannot know the contents of what remains to be analyzed. If a call to an unknown procedure *B()* appears in the code of procedure *A()*, the compiler cannot know if this is an error or if procedure *B()* is defined in the part of the file that remains to be analyzed.

```
/* Parses the WhileStm symbol                                      */
/*                                                                 */
/* WhileStm ::= <WHILE> <LPAREN> Expr <RPAREN> Statement           */
private Statement parseWhileStm (SymbolTable symtab)
{
	Token tk;
	Expression cond;
	Statement body;
	Statement stm = null;
	int [] lsync = { };
	int [] rsync = { INT, CHAR, BOOLEAN, IF, WHILE, ENDP, RBRACE };
	try
	{
		int [] expected = { WHILE };
		switch (nextToken.getKind())
		{
			case WHILE:
			{
				match(WHILE);
				tk = match(LPAREN);
				cond = parseExpr(symtab);
				match(RPAREN);
				body = parseStatement(symtab);
				{ stm = actionWhileStatement(tk,cond,body); }
				break;
			}
			default: throw new SintaxException(nextToken,expected);
		}
	}
	catch (Exception ex)
	{
		catchError(ex);
		skipTo(lsync,rsync);
	}
	return stm;
}
```

Fig. 9. A function parsing the WhileStm symbol.

The solution implemented in the URIUM compiler is to perform two passes. The first pass is performed with a parser called **UriumHeaderParser**, which is explained in Lecture04. The aim of this first pass is to obtain the declaration of the procedures defined in the parsed file, but without parsing the body of these procedures. The result of this first pass is a **LibraryDeclaration** data structure containing the list of procedures described in the file. These procedures are Procedure objects in which the body field is left empty, that is, the list of procedure instructions is not included. The main class, **UriumCompiler**, executes the **UriumHeaderParser** parser on the "*Main.ur*" file and on all files imported from it, storing the obtained **LibraryDeclaration** objects in the symbol table (**SymbolTable**). These classes are part of the **urium.ast.struct** package.

The second pass is performed with the **UriumBodyParser** parser and explained in Lecture05. In this second pass, the symbol table obtained in the first pass is used and the previously studied "*.ur*" files are re-parsed. At the beginning of the analysis of a file, the corresponding **LibraryDeclaration** object is declared as the active library within the symbol table. Each time a procedure definition is reached in the file, the corresponding **Procedure** object created in the first pass is searched for. In this second pass, the body of the procedure is parsed, obtaining the list of statements for the procedure and storing it in the body field.

The result of the two passes is a symbol table where the **LibraryDeclaration** objects contain all the information for the input URIUM files.

### *4.5. The analysis phases with an automatic tool*

Manually programming the different stages of analysis allows students to better understand the theoretical aspects of this subject, but it becomes very tedious when the grammar to be analyzed contains a large number of rules. Any modification to the grammar (to include improvements to the language) requires recalculating the lexical automaton, the prediction sets for the syntactic rules, and redoing the semantic actions. It is much more convenient to use an automated tool that generates the lexical, syntactic, and semantic analyzers from a formal specification.

Numerous tools exist for the automatic generation of parsers, such as Flex/Bison (Levine (2009)), ANTLR (Parr (2013)), CUP (Hudson *et al* (2015)), GOLD (Cook (2012), and CoCo/R (Mössenböck *et al* (2010)). In this course, JavaCC has been chosen as the parser generator because it is one of the most widely used tools among Java programmers (JavaCC Team (2026)). Lecture 06 is dedicated to introducing this tool and describing the URIUM specification in its format. The result is two new versions of the UriumHeaderParser and UriumBodyParser classes, generated automatically. During the session, modifications to the language can be proposed, and it can be seen how much easier it is to work with this type of tool. The course can be adapted to use a different tool. If there are no modifications to the Abstract Syntax Tree, the compiler's operation should be independent of the parser implementations.

## 5. THE COMPILER'S BACK-END

### *5.1. Intermediate code generation*

Once the analysis phase is complete, all the information from the source code is contained in the abstract syntax tree. The compiler must translate this information into the assembly code for the target platform. In most compilers, this translation is performed in two phases. First, the contents of the abstract syntax tree are translated into intermediate code, and then this intermediate code is translated into assembly code. The intermediate code is a pseudo-assembly that mimics typical processor instructions, but without considering architecture-specific properties such as the register set or specific instructions. In this way, the same intermediate code can be translated for different target platforms, which is known as the compiler's back-end.

One option for designing a compiler is to use intermediate code already defined by a compiler family and leverage the back-ends of those tools. For example, the GCC compiler family uses RTL as its intermediate code (Stallman *et al* (2026)), so one way to build a compiler would be to generate RTL from the source language (known as the compiler's front-end) and link it to the back-ends included in GCC. Another option, used for example in the Rust compiler, is to use LLVM (Lattner and Adve (2004)) to develop the back-end. This way, to build the compiler, one could translate to LLVM-IR (the intermediate representation of LLVM) and link it to the back-ends provided by this tool.

The goal of the URIUM compiler is to serve as a foundation for teaching a compiler design course. For this reason, the course also includes defining a custom intermediate code and programming the back-ends for different platforms. This allows students to delve much deeper into the matter and connect this knowledge with other matters such as computer architecture and operating systems.

The intermediate code defined by the URIUM compiler is a three-address code consisting of 24 instructions: labels to identify positions to reference in jumps or calls (LABEL), arithmetic operations (ASSIGN, ADD, SUB, MUL, DIV, MOD, INV), logical operations (AND, OR, NOT), jumps (JUMP, MPEQ, JMPNE, JMPGT, JMPGE, JMPLT, JMPLE, JMP1), procedure calls (PARAM, PRECALL, CALL, ENDP), and an instruction to obtain a reference to a variable (POINTER). The URIUM compiler includes the **urium.code** package with the classes dedicated to defining these instructions and the translation process.

The translation process involves examining the abstract syntax tree of each library (an object of the class **urium.ast.struct.LibraryDeclaration**) and creating an object with its equivalent intermediate code (an object of the class **urium.code.LibraryCodification**). This is done by analyzing the statements in URIUM and generating the intermediate code with the structure associated with each statement. Fig. 10 shows an example of the intermediate code structure associated with the while statement. The "*Condition* *code*" and "*Body instruction code*" blocks are generated by analyzing the loop's conditional expression and the statements in its body. In this way, a semantic tree is transformed into a list of basic instructions.

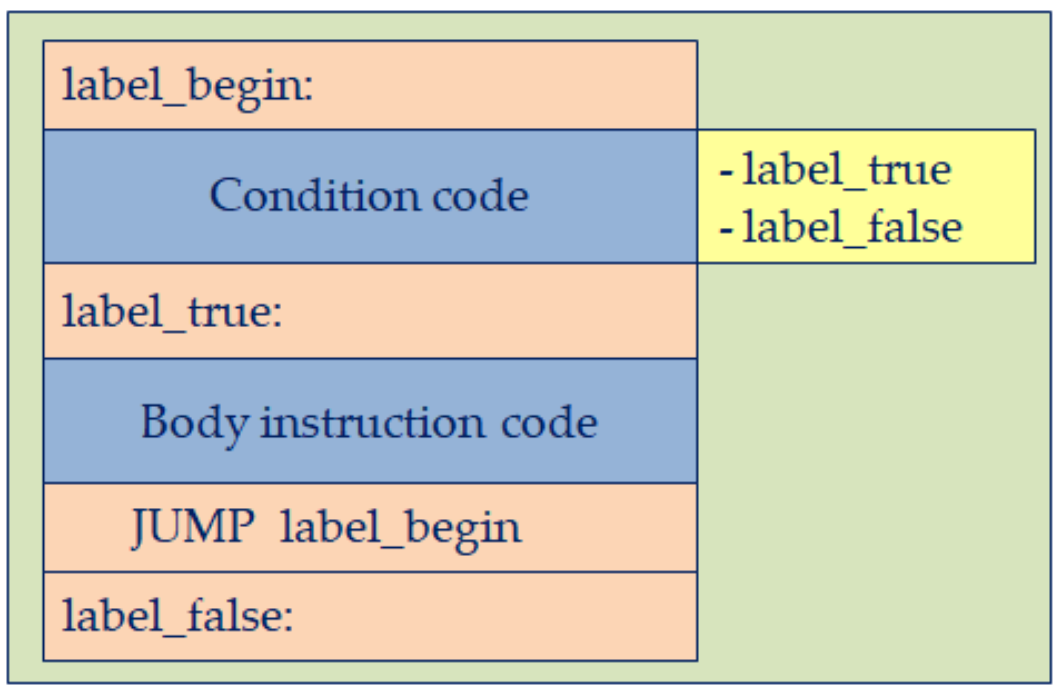

Fig. 10. Intermediate code structure that implements the WHILE statement.

Lecture 07 is dedicated to explaining the entire process of generating intermediate code for the URIUM compiler. This session explains the intermediate code defined in the compiler and the classes used to represent it, as well as the translation process using the intermediate code structures associated with each type of URIUM instruction and expression. During the lesson, students may be asked to translate new instructions or expressions (for example, the do-while statement) as an exercise.

### *5.2. Backend for MIPS32 simulator*

The first step in generating assembly code for any platform is to design the memory management at runtime. The URIUM language uses only basic data types and does not require dynamic memory. This limits memory management to the stack memory. Managing this type of memory involves defining the contents of each procedure's activation record and the call and return processes.

The first target platform included in the URIUM compiler is the MIPS32 processor (MIPS Technologies Inc. (2016)) simulated with Qt-SPIM (Larus (2023)). MIPS (Microprocessor without Interlocked Pipeline Stages) is the name of a family of RISC processors developed by MIPS Technologies (now part of GlobalFoundries). For decades, MIPS ISA-based processors have been the basic hardware of numerous workstations, game consoles, automotive systems, and networking equipment. In 2021, the company decided to abandon the MIPS architecture and focus on RISCV. However, the MIPS32 architecture remains a very good starting point for compiler development due to the simplicity and power of its instruction set. Qt-SPIM is a simulator that runs programs written in MIPS32 assembler. The tool was developed by James Larus and is freely distributed. The simulator develops a minimal set of system calls that allow access to a console and files.

The classes of the URIUM compiler dedicated to generating MIPS32 assembly code are located in the **urium.none_mips32** package. This package includes classes to describe the processor's register set and instruction set. It also includes classes to translate intermediate code into assembly code and to store the final representation of each library. The MIPS32 back-end also includes the code associated with the native libraries **urium.Console** and **urium.Program**. The final result of the compilation process is a file that combines the translations of all the libraries used in the source code, as well as common code containing the program's starting point, exception definitions, and other auxiliary definitions.

Lecture 08 of the course explains all these aspects, as well as the translation of each instruction from intermediate code to assembly code. This lesson can be divided into several sessions to explain in more detail both the MIPS processor architecture and the management of stack memory and the translation process. During these sessions, examples of the translation of various applications programmed in URIUM and their step-by-step execution in the Qt-SPIM simulator are shown (Fig.11).

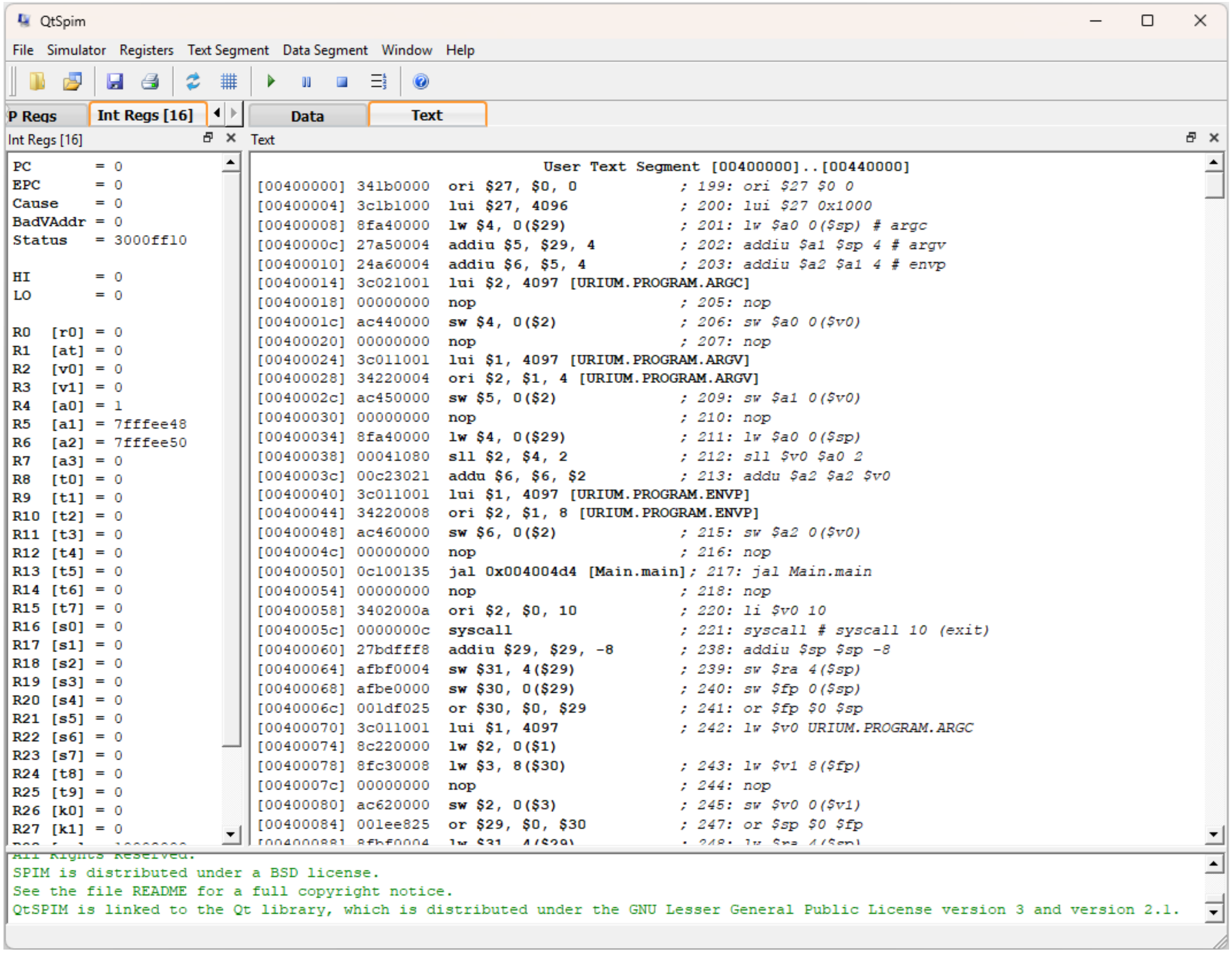

Fig. 11. Example of a Qt-SPIM execution of an application programmed in URIUM.

### *5.3. Backend for MS-Windows on Intel64*

The URIUM compiler backend for MS-Windows on Intel64 architecture (Intel Corporation (2026)) uses the 64-bit version of MASM (Microsoft Macro Assembler) (Microsoft Corporation (2024)). This version is called *ml64* and is distributed freely, for example with the Visual Studio development environment. The assembler instructions use the standard Intel syntax (other assemblers, such as GAS, use the AT&T syntax). In this syntax, register names are written without any prefix, and the instructions place the target address first, followed by the source address. For example, the instruction "**mov ebx eax**" copies the value of the EAX register to the EBX register.

Lecture 09 introduces the Intel64 architecture (the set of registers and assembler instructions required by the URIUM compiler) and explains the classes included in the compiler for developing this backend. These classes are located in the **urium.win_x64** package and include the representation of the registers and instructions of this platform, as well as the translation process between the intermediate code and the assembler code used on MASM.

In operating systems, system calls are used when a running process needs to access a resource or execute a system service. In the old DOS operating system, these system calls were translated into software interrupts using the 21h code for the **int** assembly instruction. Other operating systems running on the Intel64 architecture use the **syscall** instruction to develop system calls. MS-Windows has eliminated this form of system calls and replaced them with calls to a set of functions included in a standard library called *kernel32.lib* (Microsoft Corporation (2026)). Thus, to use system calls (for example, to read or write to the console), it is necessary to use these functions and link the code with the *kernel32.lib* library. The URIUM compiler also needs to use functions from the *shell32.lib* library to manage application arguments.

The backend for MS-Windows on Intel64 includes the assembly representation of the native libraries **urium.Console** and **urium.Program**, which uses function calls defined in *kernel32.lib* and *shell32.lib*. The compilation scripts include both the execution of the URIUM compiler and the execution of *ml64* to assemble and link the compiler output, ultimately generating an executable file for the MS-Windows console.

### *5.4. Backend for Linux on AMD64*

The AMD64 architecture (Advanced Micro Devices Inc. (2020)) is equivalent to the Intel64 architecture. In 1999, AMD announced the x86-64 architecture as a 64-bit extension of the 32-bit x86 architecture, aiming for full backward compatibility. It was first used in AMD Opteron processors in 2003. The architecture was later adopted by Intel and renamed Intel64. The URIUM compiler backend for Linux on AMD64 architecture uses the GNU assembler (Free Software Fundation (2026)). This assembler is part of the GNU Binutils package and is used as the default backend for the GNU Compiler Collection (GCC). In the Linux shell, this assembler is accessed with the *as* command. This command takes AT&T syntax assembler files as input and generates object files as output. To obtain the final executable file, the object files must be linked with the *ld* command.

The URIUM compiler backend dedicated to generating assembly code for Linux on AMD64 architecture is located in the urium.linux_amd64 package. The classes dedicated to describing the architecture are different from those used in the MS-Windows backend because the assembler syntax used is different in both backends. The compiler distribution includes the representation of the native libraries urium.Console and urium.Program. Access to operating system services is performed using the syscall assembler instruction (Linux Man-pages Project (2026)).

Lecture 10 of the course is dedicated to explaining this backend. If the MS-Windows backend has been explained previously, this lecture can focus on explaining how system calls are handled in Linux and dedicate less time to the translation process, which is very similar in both backends. The course can also be adapted to include only one of these backends and dedicate several sessions to explaining in more detail the architecture, memory management, and the process of translating intermediate code into assembly code.

### *5.5. Backend for RISC-V simulator*

RISC-V is an open-source hardware architecture that defines a RISC-like instruction set for microprocessor operation (Waterman and Asanović (2014)). The project began in 2010 at the University of California, Berkeley. RISC-V has achieved growing importance in the processor market thanks to its open, modular, and license-free nature, which has enabled rapid adoption by companies, research centers, and hardware manufacturers worldwide.

Lecture 11 of the course develops a URIUM compiler backend to generate 64-bit RISC-V processor assembler. This lesson proposes using the RARS simulator to test this code (Landers (2023)). RARS (RISC-V Assembler and Runtime Simulator) was developed by Pete Sanderson and Ken Vollmar in 2017 and is freely downloadable from its official repository. RARS develops a simulation of the RV32IMFN processor (base 32-bit instruction set + multiplication, floating point, and user-level interrupts). Version 1.6 also adds the ability to simulate the 64-bit base variant. The simulator develops a minimal set of system calls that allow, for example, access to the text console or the file system.

The code for this backend is located in the **urium.none_riscv** package. As with previous backends, this package includes classes dedicated to describing the processor register set and the assembly instruction set required by the URIUM compiler, as well as classes that handle the translation of intermediate code into RISC-V assembly code. The URIUM compiler distribution includes the translation of native libraries into RISC-V assembly. This code uses the system calls included in RARS to access the console (Fig. 12).

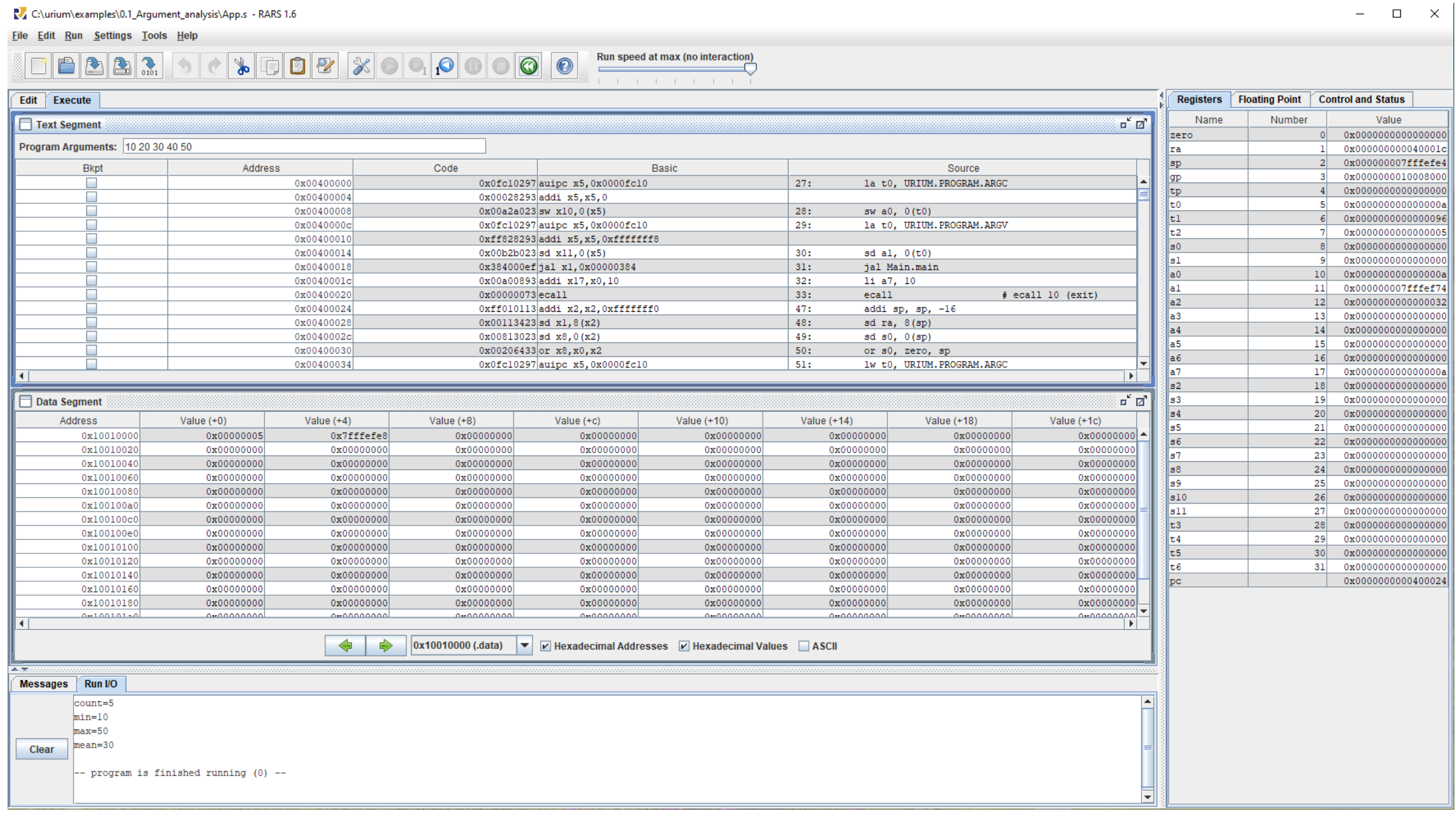

Fig. 12. Example of a RARS execution of an application programmed in URIUM.

## *5.6. Backend for Linux on RISC-V*

The definition of the RISC-V architecture as open hardware allows manufacturers to develop their products without licensing costs. One of the companies dedicated to the design and manufacture of devices based on the RISC-V architecture is StarFive. One of the company's products is the VisionFive2 development board (StarFive Technology (2022)), which we will use as a demonstrator of the URIUM compiler on a real RISC-V platform. Since 2023, the Debian community has officially supported its Linux distribution on the VisionFive2 board (Fig. 13).

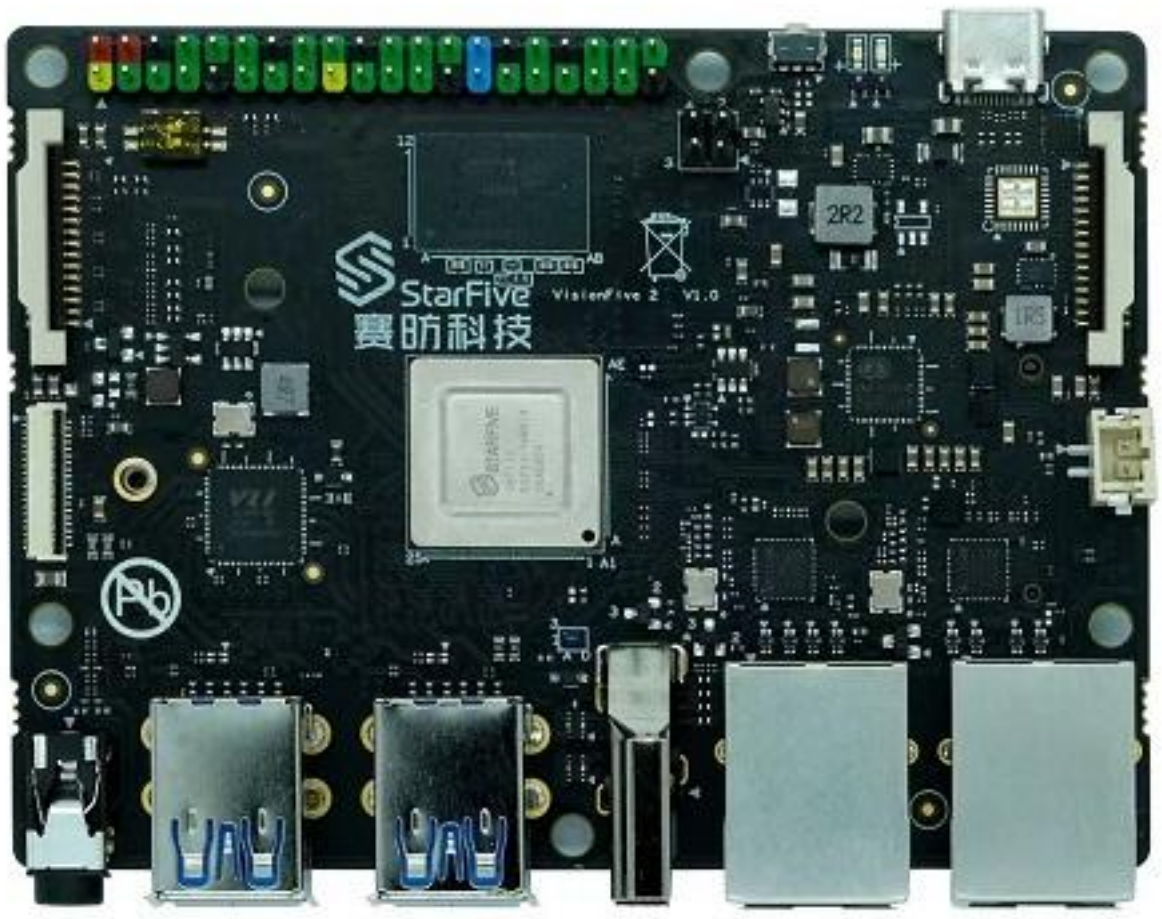

Fig. 13. VisionFive2 development board used as RISC-V platform.

Lecture 12 is dedicated to explaining a backend of the URIUM compiler that generates Linux assembler for the RISC-V processor. In this case, the assembler associated with the URIUM source code is the same as that presented in Lecture 11, so the classes dedicated to this backend are those included in the urium.none_riscv package. To adapt this code to the Linux platform, changes are made to the assembler code of the native libraries (**urium.Console** and **urium.Program**) and to the assembly and linking process (which in this case uses the Linux commands *as* and *ld*). If Lecture 11has been included in the course (and, therefore, time has been dedicated to explaining the RISC-V architecture), this final lesson can dedicate more time to explaining Linux system calls and testing the different examples included in the distribution.

## 6. ADAPTATIONS OF THE COURSE

This syllabus is designed for a typical semester-long course. To achieve this, some lessons can be divided into several sessions. The course can be adapted by reducing the number of backends explained and dedicating more time to lessons related to the analysis phase. The proposed content corresponds to the practical part of the course. It is assumed that these sessions should be accompanied by theoretical sessions describing the fundamental principles underlying each stage of the compiler. Numerous manuals already exist that thoroughly cover this theoretical content, for example Aho *et al* (2006), Appel and Palsberg (2002), Cooper and Torczon (2011), Grune et al (2012), Louden (2007), or Wirth (2017). This course focuses on how to apply theoretical knowledge to a real-world program. The syntactic/semantic analysis chosen for developing the compiler is a top-down parsing method. The course can be modified to include a bottom-up parsing method (using the SLR or LALR algorithms), but manually programming these parsers becomes more complex.

## 7. EVALUATION PROJECTS

The specification of the URIUM language used in the course is very simple and allows for different projects to be proposed as a way to evaluate the course. Some of these projects could be the following:

- Add the do-while and for loops to the language.
- Add the switch-case-default statement to the language.
- Add bitwise operators (shifts, complement, AND, and OR) to the language.
- Add compound assignment operators (+=, -=, *=, /=, %=) to the language.
- Add the byte and short data types.
- Add the float data type.
- Add the double data type.
- Add the long data type (on 64-bit architectures).
- Add code optimization algorithms.
- Add a backend for the ARM architecture.

## 8. FURTHER EXPLORATION

This course builds a fully functional compiler from scratch. To accomplish this in a one-semester course, the programming language used must be simple in both its instruction set and supported data types. A more complex language requires greater effort and cannot be covered in a course of this duration. However, the work developed here can serve as a starting point for a more advanced compiler design course.

Some aspects to consider in an advanced course would be the following:

- Adding complex data types (structures, arrays, unions)
- Adding code optimization and register allocation algorithms
- Adding dynamic memory management
- Adding garbage collection
- Adding object-oriented programming
- Adding exception handling
- Adding functional programming

Any case, the aim of this advanced course would be not only to explain the theoretical foundations of these features but also the practical programming aspects to incorporate them into the URIUM language and its compiler.

## References

Advanced Micro Devices, Inc. (2020). AMD64 Technology: AMD64 Architecture Programmer's Manual Volumes 1-5. Revision 4.00, Publication No. 40332.

Aho, A.V., Lam, M.S., Sethi, R., & Ullman, J.D. (2006). Compilers: Principles, Techniques, and Tools (2nd ed.). Addison-Wesley.

Aiken, A.. (1996). Cool: A portable project for teaching compiler construction. In ACM SIGPLAN Notices, Vol. 31. 19–24. https://doi.org/10.1145/381841.381847

Appel, A.W., & Palsberg, J. (2002). Modern Compiler Implementation in Java (2nd ed.). Cambridge University Press, Cambridge, UK.

Cook, D.. (2012). The GOLD Parsing System. Available at http://www.goldparser.org/.

Cooper, K., & Torczon, L. (2011). Engineering a Compiler (2nd ed.). Morgan Kaufmann.

Free Software Foundation. (2026). Using as: The GNU Assembler. GNU Project. Available at https://sourceware.org.

Grune, D., van Reeuwijk, K., Bal, H.E., Jacobs, C.J.H., & Langendoen, K.G.. (2012). Modern Compiler Design (2nd ed.). Springer, New York, USA. https://doi.org/10.1007/978-1-4614-4699-6

Hudson, S.E., Ananian, C.S., Flannery, F., Wang, D., Appel, A.W., & Petter, M. (2015). CUP LALR Parser Generator for Java. Technical University of Munich.

Intel Corporation (2026). Intel® 64 and IA-32 Architectures Software Developer's Manual. Intel Corporation, Santa Clara, CA, USA. https://www.intel.com/content/www/us/en/developer/articles/technical/intel-sdm.html Combined Volumes: 1, 2A, 2B, 2C, 2D, 3A, 3B, 3C, 3D, and 4.

JavaCC Team. (2026). JavaCC: The Java Compiler Compiler. Available at https://github.com/javacc/javacc

Landers, B. (2023). RARS: RISC-V Assembler and Runtime Simulator. Available at https://github.com/TheThirdOne/rars

Larus, J.R. (2023). QtSpim: A MIPS32 Simulator. Available at https://spimsimulator.sourceforge.net/ (Version 9.1.22).

Lattner, C., & Adve, V. (2004). LLVM: A Compilation Framework for Lifelong Program Analysis and Transformation. In Proceedings of the international symposium on Code generation and optimization (CGO). 75–88.

Levine, J. (2009). Flex & Bison: Text Processing Tools. O'Reilly Media, Inc.

Linux Man-pages Project. (2026). fork(2) — Linux manual page. Linux Kernel Organization. Available at https://man7.org (Section 2: System Calls).

Louden, K.C. (2007). Compiler Construction: Principles and Practice. Thomson Learning.

Microsoft Corporation. (2024). Microsoft Macro Assembler (MASM). Software utility. https://learn.microsoft.com/en-us/cpp/assembler/masm/microsoft-macro-assembler-reference (Version 14.x / ML and ML64).

Microsoft Corporation. (2026). Programming reference for the Win32 API. https://learn.microsoft.com/en-us/windows/win32/api/

MIPS Technologies, Inc. (2016). MIPS32® Architecture for Programmers, Volume II-A: The MIPS32® Instruction Set Manual (revision 6.06 ed.). MIPS Technologies, Inc., Sunnyvale, CA. Document Number: MD00086.

Mössenböck, H., Löberbauer, M., & Wöß, A. (2010). Coco/R: A Compiler Generator. Institute for System Software, Johannes Kepler University Linz. Available at https://ssw.jku.at/Research/Projects/Coco/

Parr, T. (2013). The Definitive ANTLR 4 Reference (2nd ed.). Pragmatic Bookshelf.

Roberts, E.. (2001). An Overview of MiniJava. In Proceedings of the 32nd SIGCSE Technical Symposium on Computer Science Education. ACM, 1–5. https://doi.org/10.1145/364447.364525

Stallman, R.M., & GCC Developer Community. (2026). GNU Compiler Collection (GCC) Internals. Free Software Foundation. https://gnu.org/ (Chapter: Register Transfer Language RTL).

StarFive Technology. (2022). VisionFive 2: The World's First RISC-V Single Board Computer with an Integrated 3D GPU. https://starfivetech.com.

Waterman, A., & Asanović, K.. (2014). The RISC-V Instruction Set Architecture. Technical Report UCB/EECS-2014-54. University of California, Berkeley. Available at https://www2.eecs.berkeley.edu/

Wirth, N. (1976). Algorithms + Data Structures = Programs. Prentice-Hall, Englewood Cliffs, NJ.

Wirth, N. (2017). Compiler Construction (revised edition ed.). ETH Zürich, Zürich, Switzerland. https://people.inf.ethz.ch/wirth/CompilerConstruction/CompilerConstruction1.pdf (Slightly revised version of the 1996 Addison-Wesley edition).